\documentclass[11pt,a4paper]{JHEP3}
\usepackage{yfonts, amsmath,epsfig}
\def\dd{\mathrm{d}}
\def\bx{\mathbf{x}}
\def\bo{\mathbf{0}}
\def\bq{\mathbf{q}}
\def\NN{\mathcal{N}}
\def\co{\mathcal{O}}
\def\cp{\mathcal{\,P\!\!\!}}
\def\Dis{\tilde d}
\def\mass{m_q}
\def\re{{\mathrm{Re}}}
\def\im{{\mathrm{Im}}}

\def\be{\begin{equation}}
\def\ee{\end{equation}}
\def\beqa{\begin{eqnarray}}
\def\eeqa{\end{eqnarray}}

\preprint{ITP-UU-10/38}

\title{
Sum rules, plasma frequencies and Hall phenomenology in holographic plasmas\\
}
\author{Javier Mas\footnote{jamas@fpaxp1.usc.es}, Jonathan P. Shock\footnote{shock@fpaxp1.usc.es}  and Javier Tarr\'\i o\footnote{l.j.tarriobarreiro@uu.nl}
\vspace{0.75cm}\\
Departamento de F\'\i sica de Part\'\i culas,
Universidade de Santiago de Compostela 
and
Instituto Galego de F\'\i sica de Altas Enerx\'\i as (IGFAE)\\
E-15782 Santiago
de Compostela, Spain
\vspace{0.35cm}\\
Kavli Institute for Theoretical Physics China, CAS\\
Beijing 100190, China
\vspace{0.35cm}\\
Institute for Theoretical Physics, Universiteit Utrecht,\\
3584 CE, Utrecht, The Netherlands
}
 
\abstract{
We study the AC optical and Hall conductivities of $Dp$/$Dq$-brane intersections in the probe approximation and use sum-rules to study various associated transport coefficients. We determine that the presence of massive fundamental matter, as compared to massless fundamental matter described holographically by a theory with no dimensional defects, reduces the plasma frequency. We further show that this is not the case when the brane intersections include defects. We discuss in detail how to implement correctly the regularization of retarded Green's functions so that the dispersion relations are satisfied and the low energy behaviour of the system is physically realistic.
}

\begin{document}

\section{Summary of results}

The interpretation of the physics encountered in the Quark-Gluon Plasma (QGP) from the point of view of the gauge/gravity correspondence finds an overwhelming support in the calculation of thermodynamic quantities and hydrodynamic transport coefficients, which are essential to understand the late-time behavior of the QGP in computational simulations. Thermodynamic quantities in the field theory are calculated by evaluating the classical action of the dual gravitational model on-shell. In contrast, hydrodynamic quantities are obtained by calculating the correlation functions, which in lattice simulations rely on postulating the form of the spectral density. As pointed out in \cite{Romatschke:2009ng}, sum rules may be used as constraints on the spectral densities obtained from the gauge/gravity correspondence, relating the integral of --roughly-- the spectral density with thermodynamic quantities and transport coefficients. 


In this paper we study sum rules for the current-current correlator of fundamental matter at finite temperature. A similar analysis has been done for $\NN=4$SYM in \cite{Baier:2009zy}, where a considerable difference between the calculations at strong and weak coupling was found. We will restrict to the  strongly coupled regime, using the AdS/CFT correspondence to study different plasmas described in the string theory side by intersections of two stacks of $Dp$-branes with different dimensionalities. We will consider $N_f$ probe $Dq$-branes in the background created by $N_c\gg N_f$ $Dp$-branes.

The first result of the present work is the determination of the effects of having matter transforming in the fundamental representation on the collective modes of the plasma, namely on the plasma frequency. This is related to the current-current correlator  by a sum rule for the AC conductivity \cite{Nozieres:1958zza}
\be
\int_{-\infty}^\infty \re\, \sigma(\omega) \dd \omega = \epsilon_0 \, \omega_p^2 \, .\label{plasmafreq}
\ee
 The AC conductivity is related to the retarded current-current correlator, $\tilde G^R(\omega)$, by the expression $i\omega \sigma(\omega)= \tilde G^R(\omega)$. We can then use the prescription in \cite{hep-th/0205051}  to obtain the retarded Green's function for the holographic system and evaluate the integration of the sum rule. This will give us a divergent retarded correlator (and conductivity) in the large frequency limit. To treat this divergence we have to take a subtraction scheme in which we take the difference between two of these correlators. This is analogous to the Ferrell-Glover-Tinkham sum rule in superconductivity studies, which relates the difference between Green's functions at different temperatures (see for example \cite{FGTex}). In this case the evaluation of the sum rule for the conductivity will not give us the value of the plasma frequency but the difference between $\omega_p^2$ in both setups. Schematically, we find
\be\label{firstres}
\epsilon_0\, \omega_p^2(M_q,T)- \epsilon_0\, \omega_p^2(M_q=0,T=0) \propto \begin{cases} 
- N_f N_c\, M_q^2/\lambda & \text{if no defect} \\
 0 & \text{if defect}
\end{cases}\, ,
\ee
with $M_q$ the dimensionful mass of the fundamental matter and $\lambda$ the 't Hooft coupling\footnote{In  $p\neq 3$, $\epsilon_0$ is dimensionful. Our result \eqref{firstres} suggests that it is actually $\sim 1/\lambda$.}. Our results show that the value of the plasma frequency \emph{gets reduced} in a thermal system when massive matter in the fundamental of $SU(N_c)$ is included in a brane intersection without defects, the reduction being proportional to the mass of the fundamental degrees of freedom squared. However, we are not aware of a calculation of the plasma frequency for a stack of $Dp$-branes, therefore we are calculating the effects of fundamental matter on an unknown quantity. It presumably scales with $N_c^2$, whereas our correction scales with $N_f N_c\ll N_c^2$ and thus is subleading in the probe approximation. 

The impact of defects\footnote{i.e., theories where the fundamental matter lives only in a subspace of the original gauge theory.}  in the brane intersection in the system is noteworthy. Our results suggest that the integration of the spectral function over the frequency in the transverse vector channel (\emph{i.e.}, the integration of the real part of the conductivity) is zero or non-zero depending on whether the theory is a defect or not. This suggests that this is an example of an observable which is sensitive not only to the number of dimensions of the plasma in the IR, but also to the dimensionality of the UV theory (which can be reduced by means of a Kaluza-Klein compactification). Though we have not been able to find a good reason for this difference, we have checked that this rule holds true for a large class of brane intersections.  

We turn our attention next to a sum rule involving the Hall conductivity. To study it we need to include in our setup an external magnetic field as well as a finite baryon density. Focusing on the $D3$/$D7$ intersection in the probe approximation, we will consider the Hall angle, defined as the quotient between the Hall and optical conductivities $t_H=\tan \theta_H\equiv \sigma_{xy}/\sigma_{xx}$, which gives the angle between the electric current and the electric field. Following \cite{drewcoleman} we numerically determine the conductivity tensor as a function of the frequency and integrate $t_H(\omega)$ in the whole frequency range (technically we just integrate up to a cutoff, however the remaining contribution can be made arbitrarily small by improving the numerics).

We have to deal with a holographic system which couples the modes of interest, due to the presence of the magnetic field, which breaks parity invariance. Going to a circular polarization in the transverse plane to the magnetic field we can study the system in a maximally coupled basis\footnote{Maximally coupled at the level of the action which leads to a completely decoupled set of equations of motion.} which turns out to be simpler, and obtain the Hall and optical conductivities once the solution in the circular polarization basis is known.

This allows us to calculate the Hall frequency, $\omega_H$, by means of another sum rule\footnote{Here $t_H=\tilde G_{xy}/\tilde G_{xx}$ is dimensionless. Thus, this equation for the Hall frequency, which is derived in  \eqref{eq:hallfreqdef}, yields a linear result for $\omega$ , in contrast to the case for the plasma frequency, $\omega_p$, in (\ref{plasmafreq}).}
\be
\omega_H =  \frac{1}{\pi} \cp \int_{-\infty}^\infty  \re \,t_H(\omega)\dd \omega \, .
\ee
The Hall frequency can be understood as the frequency at which the phenomenology due to the Hall effect dominates over the diffusive processes in a non-ideal fluid. We present the numerical determination of $\omega_H$ in figure \ref{fig:hallfreq}. At low masses of the charge carriers we observe that the Hall frequency is increasing, meaning that the diffusive processes become more important than the Hall effect for the medium frequencies of the signal. At a given mass value (depending on both the magnetic field and the density of charge carriers) the Hall frequency attains a maximum value, and for larger masses $\omega_H$ decays with a power of the mass.

\subsubsection*{Outline of the paper}

In section \ref{sec:general} we introduce some general considerations about the retarded Green's function, which will lead to a set of general sum rules. Concretely, we will focus on sum rules for the conductivity.

In section \ref{sec:effects} we investigate the consequences of the conductivity sum rule in the case of a $Dp$/$D(p+4)$ intersection. We reason that the satisfaction of the sum rules in these cases implies a reduction of the plasma frequency. We continue the discussion to more general brane intersections in section \ref{sec:defects}.

In section \ref{sec:hall} we turn our attention to a sum rule satisfied by the quotient between the Hall and optic conductivities (the Hall angle), and we use it to estimate the Hall frequency as defined in the text.

Lastly, in section \ref{sec:conclu} we write some conclusions and future directions.

\section{Sum rules from causal considerations}\label{sec:general}

The retarded correlator 
\be\label{eq:GRet}
G^{R}(t,\bx;t',\bx') = -i \Theta(t-t') \left< [\co(t,\bx),\co(t',\bx')] \right>_0 \equiv - i \Theta(t-t') \rho(t,\bx;t',\bx')\, , 
\ee
plays an important r\^ole in linear response theory, since it determines the expectation value of the operator $\co$ in the presence of an infinitesimal source $\delta$
\be
\left< \co (t,\bx)\right>_\delta = i \int \dd t'\dd \bx' G^R(t,\bx;t',\bx') \delta(t',\bx') \, .
\ee
In this paper we are interested in the case where the source takes the form of a perturbative electric field $A_\nu(t')$. The expectation value for the current then takes the form
\be\label{eq:LRT}
\left< J^\mu (t,\bx)\right>_A = \int \dd t'\dd \bx' \Theta(t-t') \left< [J^\mu(t,\bx),J^\nu(t',\bx')] \right>_0 A_\nu(t',\bx') \, ,
\ee
thus the Green's function of interest will be the retarded current current correlator.

\subsection{Dispersion relations}\label{sec:disprels}

The presence of the Heaviside theta function in the definition of the retarded correlator ensures causality in the effect of the source, avoiding signals propagating into the backward lightcone. For this reason the correlator naturally entering in linear response theory is the retarded one, since it guarantees that no signal precedes the perturbation. It also implies that the Fourier transformed correlator $\tilde G(\omega, \bq)$ satisfies the Kramers Kronig (dispersion) relations (see for example \cite{Landau})
\beqa\label{eq:disprel1}
 \re\, \tilde G_\Delta(\omega,\bq) -\tilde G^\infty_{\Delta} & = &  - \frac{1}{\pi} \cp \int_{-\infty}^\infty \frac{\im\, \tilde G_\Delta(\mu,\bq)}{\omega-\mu}\dd \mu\, , \\
\label{eq:disprel2}
\im\, \tilde G_\Delta(\omega,\bq) & = &  \frac{1}{\pi} \cp \int_{-\infty}^\infty \frac{\re\, \tilde G_\Delta(\mu,\bq)-\tilde G^\infty_{\Delta}}{\omega-\mu}\dd \mu\, ,
\eeqa
where ${\mathcal{P}}$  is the Cauchy principal value around the pole at $\mu=\omega$. 

The Green's function usually presents divergences in the large frequency limit, and therefore the results \eqref{eq:disprel1} and \eqref{eq:disprel2} may be ill-defined. This can be fixed by subtracting the divergent behavior in the $|\omega|\to\infty$ limit. In general one should choose a scheme to perform this regularization. One usual choice in holographic contexts is to take off the zero temperature Green's function, given by powers and logarithms of the frequency  ($\tilde{G}_\Delta=\tilde{G}_T-\tilde{G}_{T=0}$). It may be that the exact expression for $\tilde{G}_{T=0}$ is not known, in which case the comparison should be carried out with a specific scheme, and the interpretation of the results has to take this scheme into consideration. It turns out that for physical quantities we can show that there is no scheme dependence.

Therefore, in order to apply \eqref{eq:disprel1} and \eqref{eq:disprel2}, we have to make sure that we have subtracted all the divergent contribution to the correlator at large frequencies \cite{Romatschke:2009ng,CaronHuot:2009ns}. We have marked this in the previous equation by the ${}_\Delta$ subscript.

It may be that although the subtracted function $\tilde{G}_\Delta$ has had all UV divergences removed, a constant piece may still survive ({\it ie.} $\tilde{G}_\Delta(\omega\rightarrow \infty)\ne 0$) \cite{Romatschke:2009ng} and thus the integral in the Kramers Kronig relation diverges. This constant (real) piece, $\tilde{G}_\Delta^\infty$, which remains in the UV, corresponds to a contact term given by a $\delta(t)$ factor in the two-point correlator function, and is not strictly acausal. This factor may be obtained by OPE techniques, relating it to the expectation values of local operators \cite{CaronHuot:2009ns}. The Kramers Kronig relation then holds for the fully subtracted Green's function which has had the contact term removed.

Causality of the Green's function, as dictated by the Kramers Kronig relation, determines the symmetry properties of the spectral function. In general, the spectral function of bosonic hermitian operators $\tilde \rho(\omega,\bq)$ is an odd function of the frequency defined as $\tilde \rho_\Delta= - 2 \im \,\tilde G_\Delta$, where $\tilde \rho_\Delta$ stands for the part of the spectral function that survives after the subtraction (NB. The imaginary part of the Green's function will not receive any contact term contributions, which effect solely the real part).

\subsection{Sum rules}

It is possible to extract a number of relations between integrals in the whole frequency regime and some physically relevant quantities. These relations follow from \eqref{eq:disprel1} and \eqref{eq:disprel2} and involve on the one hand an integration over the entire positive and negative frequency domain, capturing the effects occurring at infinitely small (positive and negative) times and on the other, they imply a certain behavior over the whole frequency range, including for null frequency, thus giving information about the response of the system at large times.

Recall that we have already subtracted the large frequency behavior in the integrand to make certain that these integrals converge. A stricter method would be to introduce a cutoff frequency, $\omega_\Lambda$, that effectively neglects instantaneous effects in the response of the system. In fact, in this paper we will perform the integrations numerically, such that this cutoff frequency appears naturally in the calculation. Nevertheless, the dependence of the results on the position of the cutoff (once we have subtracted the non-convergent part) gets suppressed as $\omega_\Lambda\to \infty$, and the difference between cutting the integrals off or performing the full integration becomes negligible.

The information we gain at null frequency (\emph{i.e.}, infinitely large times) has to be interpreted with some care. Here, we will restrict our studies to the probe approximation, in which we insert a probe in a never-changing background. Physically, neglecting the backreaction of the probe on the background is only valid at times lower than a specific timescale, after which the effects of the probe on the system must be taken into account (\emph{i.e.}, the time at which the transfer of momentum from the $N_fN_c$ degrees of freedom to the $N_c^2$ degrees of freedom becomes important). Therefore, we should interpret our results as valid for time periods well beyond when the perturbation is turned on, but not as large as for the probe approximation to fail.

\subsubsection*{Thermodynamic sum rule}

Perhaps the simplest sum rule we may obtain from the dispersion relations at zero momentum\footnote{For the rest of this paper we restrict to the $\bq=\bo$ case.} is the value of the retarded correlator at zero frequency. Inserting $\tilde \rho_\Delta= - 2 \im \,\tilde G_\Delta$ in \eqref{eq:disprel1}  we obtain
\be\label{eq:sumrule}
\re\,\tilde G_\Delta(0,0) - \tilde G^\infty_\Delta = - \frac{1}{2\pi} \cp \int_{-\infty}^\infty \frac{\tilde \rho_\Delta(\omega,0)}{\omega}\dd \omega\, ,
\ee
which is a thermodynamic sum rule for the spectral function. This is equation (11) in \cite{Baier:2009zy}, with a relative factor of $2\pi$ in the definition of the spectral function, and a global sign coming from the definition of the Green's function (see also \cite{Romatschke:2009ng}). To obtain this relation we used only the general definition \eqref{eq:GRet}, so it does not depend on any details of the model under consideration.

In \cite{Romatschke:2009ng} certain sum rules for the spectral density in hot gauge theories were derived. Specifically, for the ${\cal N}=4$ SYM at infinite 't Hooft coupling we have the following sum rules involving the $xy,xy$ graviton correlator (following our conventions, which differ from the ones in \cite{Romatschke:2009ng} by a factor of $2$ in the definition of the spectral function)
\be\label{eq:RomSon}
\re\, \tilde G_{\Delta}^{xy,xy} (0,0) - \tilde G^\infty_\Delta = \frac{1}{\pi} \int_0^\infty \frac{\dd \omega}{\omega}\tilde\rho^{xy,xy}_\Delta(\omega) \,,
\ee
where $\re\, \tilde G_{\Delta}^{xy,xy} (0,0)=P$ is the pressure (from hydrodynamics, see \cite{Baier:2007ix}) and $ \tilde G^\infty_\Delta = \frac{11}{5} P$ is a constant value that survives at infinitely large frequencies \cite{Romatschke:2009ng}. Subtracting and using the conformal equation of state, the l.h.s. in \eqref{eq:RomSon} is $-\frac{2}{5} \epsilon$, where $\epsilon$ is the energy density.

In \cite{Baier:2009zy} a sum rule was obtained for the R-current correlator at vanishing momentum in ${\cal N}=4$ SYM at infinite 't Hooft coupling
\be\label{eq:baier}
\re\,  \tilde G^{R-charge}_\Delta(0) -  \tilde G^{\infty}_\Delta=  \frac{1}{\pi} \int_0^\infty \frac{\tilde \rho^{R-charge}_\Delta(\omega)}{\omega}\dd \omega= 0 \, ,
\ee
where $\tilde G^\infty_\Delta=0$ and $\re\,  \tilde G^{R-charge}_\Delta(0)=0$ is a result dictated by hydrodynamics. Other examples of such sum rules  can be found in \cite{Meyer:2010ii,Springer:2010mf}.

Recently, the authors of \cite{Gulotta:2010cu} generalized these sum rules by studying the constraints of gravity on the retarded correlator. Our interest in this paper is to use the sum rules to study the behaviour of transport coefficients, since the systems we study are known to have a meromorphic retarded Green's function with causal structure.

\subsubsection*{Conductivity sum rule}

From \eqref{eq:LRT}, and using Ohm's law, we can express the AC conductivity in terms of the subtracted retarded correlator as
\be \label{eq:kubocond}
\tilde G_\Delta(\omega) = i \omega \sigma (\omega) \, .
\ee
Therefore, the pole structure of the conductivity is basically the same as in the retarded correlator, and the dispersion relations \eqref{eq:disprel1} and \eqref{eq:disprel2} have to hold for $\sigma(\omega)$. There is the possibility of an additional pole (accompanied by a delta function) at zero frequency in the conductivity. However, this pole does not affect the satisfaction of the dispersion relations discussed above. Consistency implies that the cutoff-dependent integrated conductivity
\be\label{eq:specweight}
W(\omega_\Lambda) \equiv \frac{1}{\pi} \int_{-\omega_\Lambda}^{\omega_\Lambda} \sigma(\mu)\, \dd \mu \, ,
\ee
goes to $-G^\infty_\Delta$ in the limit $\omega_\Lambda\to \infty$. To see this, notice that from \eqref{eq:kubocond} $\re\, \tilde G^R(\omega) = - \omega \im\, \sigma (\omega)$ and $\im\, \tilde G^R(\omega) =  \omega \re\, \sigma (\omega)$. Plugging this into \eqref{eq:disprel1} we obtain
\be
- \omega \im\,  \sigma (\omega) -\tilde G^\infty_\Delta=  - \frac{1}{\pi} \cp \int_{-\infty}^\infty \frac{\mu\, \re\, \sigma(\mu)}{\omega-\mu}\dd \mu\, ,
\ee
and from \eqref{eq:disprel2} for $\sigma$
\be
- \omega \im\,  \sigma (\omega) =  - \frac{\omega}{\pi} \cp \int_{-\infty}^\infty \frac{ \re\, \sigma(\mu)-\sigma^\infty}{\omega-\mu}\dd \mu\, .
\ee

As the imaginary part of the correlator goes to zero at large frequencies, \emph{i.e.}, $G^\infty_\Delta$ is real, then $\sigma^\infty=0$ follows. Assuming now that in the principal value all the limits are taken at the same rate, we can join the former integrands to obtain\footnote{Note that plugging the relation between $\re \,\sigma$ and $\tilde \rho_\Delta(\omega)$  in \eqref{eq:specweigdem} we do not recover \eqref{eq:sumrule} with $\re\,\tilde G_\Delta(0)=0$ because the Cauchy principal value to evaluate the pole at $\omega=0$ has disappeared. In particular, the r.h.s. of equation \eqref{eq:sumrule} is insensitive to delta functions.}
\be\label{eq:specweigdem}
-\tilde G^\infty_\Delta = - \frac{1}{\pi} \cp \int_{-\infty}^\infty \frac{ \re\, \sigma(\mu)}{\omega-\mu}(\mu-\omega)\dd \mu = \frac{1}{\pi} \int_{-\infty}^\infty  \re\, \sigma(\mu) \dd \mu \, .
\ee
Now, as $\re\, \tilde G^R$ is even in frequencies, the corresponding quantity $\im\, \sigma$ will be odd, meaning that the integrated conductivity \eqref{eq:specweight} is real. Therefore
\be
\lim_{\omega_\Lambda\to\infty} W(\omega_\Lambda) = -\tilde G^\infty_\Delta \, .
\ee
The integral over the whole range of frequencies is employed in the analysis of plasmas as a way to obtain the plasma frequency, giving the energy of the collective excitations in the plasma, by using the expression \cite{Nozieres:1958zza}
\be\label{eq:plasmon}
\lim_{\omega_\Lambda\to\infty} W(\omega_\Lambda) = \epsilon_0\, \omega_p^2 \, .
\ee
Because in our case we are defining the conductivity from a subtracted correlator, what we are calculating is a specific contribution to the plasma frequency (in fact it is a difference in contributions as we shall discuss shortly). Notice that we are not demonstrating here that the integral of the conductivity in the frequency domain determines the value of the plasma frequency. We are using this as a definition of the plasma frequency to calculate it in a holographic setup, instead.

It seems important at this point to summarize our results. The causal structure of the conductivity allowed us to write down equation \eqref{eq:disprel1}. For $\omega=0$ this relation gives us the difference between the Green's function at infinite frequency and at zero frequency. On top of this we use the fact that the Green's function at zero must vanish so that the real part of the conductivity does not develop a delta function at the origin. A delta function  in $Re[\sigma(\omega)]=\pi\delta(\omega)$ is not a sign of superconductivity,  but is
a universal property of systems with translational invariance  (see for example \cite{Hartnoll:2008kx}). In  systems where the quenched flavors are modelled with probes, the huge mismatch in the number of degrees of freedom $N_c>N_f$ mimics the presence of fixed scattering centers in the bulk, leading to a finite DC conductivity. In reality this is  a transient effect for times of order $N_f/N_c$ (see  \cite{Karch:2008uy}). Of course, in a setup where backreaction of flavor is taken into account, full translational invariance is recovered, and a delta function should be present. It would be  worth studying this effect,  for example, in the recently constructed model in \cite{Bigazzi:2011it}. For us here,  the absence of a delta function is a physical constraint on top of causality. This definition then gives us a way to calculate not only the difference between the zero and infinite frequency behaviours, but the absolute value of the Green's function at infinity. This itself has a physical interpretation as the plasma frequency. In a metal the plasma frequency describes the frequency of vibration of electrons about the ions, however in our system there is clearly nothing analogous to the ions. A possible interpretation of the plasma frequency may then be given by the rate of pair production by an infinitesimal electric field which causes pairs of charged particles to be pulled out of the vacuum before annihilating again. The fact that the plasma frequency is dependent only on the quark mass and not the baryon density suggests that it's a vacuum effect and not related to the excess of charge carriers.

Expression \eqref{eq:specweight} relies on the relation \eqref{eq:kubocond} and the causal structure of the conductivity (and thus the retarded correlator). See however \cite{Gulotta:2010cu} for a discussion deriving similar relations from the gravitational point of view. Equation \eqref{eq:specweight}  can also be used to obtain numerically the weight of the factors multiplying delta functions in the conductivities. These delta functions can occur in metals not presenting superconducting behavior, but that are not able to dissipate momentum. In this case, impurities will destroy the delta function giving rise to a Drude peak. In superconducting systems this is not the case, the delta functions will still be present in the presence of impurities, but their weight may get reduced by the presence of the extra scattering centers.

\subsubsection*{Hall angle sum rule}

When turning on a magnetic field the Green's function of the transverse vector channel has to be treated as a matrix, since the parity-breaking associated to the finite magnetic field induces a coupling between the fluctuations along the transverse directions to it. In this case, the Kramers Kronig relations are written as
\be
\tilde G_\Delta^H(\omega) - \tilde G_\Delta^\infty = \frac{i}{\pi} \int_{-\infty}^\infty \frac{\tilde G_\Delta^A(\mu)}{\omega-\mu} \dd \mu \, , \quad \tilde G_\Delta^A(\omega) = \frac{i}{\pi} \int_{-\infty}^\infty \frac{\tilde G_\Delta^H(\mu) - \tilde G_\Delta^\infty}{\omega-\mu} \dd \mu \, ,
\ee
where the superindices ${}^{H,A}$ denote hermitian and antihermitian parts, respectively. In this case 
$ \tilde G_\Delta^\infty$ is a constant hermitian matrix.

As we will see later, the presence of a magnetic field in the homogeneous system induces a form for the Green's function matrix (notice that now we are not subtraction the asymptotic behaviour)
\be
\tilde G = \begin{pmatrix} \tilde G_{xx} & \tilde G_{xy} \\ -\tilde G_{xy} & \tilde G_{xx} \end{pmatrix} \, .
\ee
We will focus our interest on a sum rule satisfied by the quotient $t_H(\omega)=\tilde G_{xy}/\tilde G_{xx}$. Let us sketch the procedure following \cite{Landau, drewcoleman}.

We would like to show that $t_H$ is a causal signal. If this is the case then $t_H$ cannot have poles in the upper half plane. Now, the $x-y$ component of the Green's function matrix has to be a causal function too, therefore potential acausal poles in $t_H$ must come from zeros of the diagonal component in the upper half plane. However, $\tilde G_{xx}(\omega)$ is a meromorphic function constrained to satisfy $\tilde G_{xx}(\omega)^*=\tilde G_{xx}(-\omega^*)$. In this case, complex variable analysis leads to the conclusion that $\tilde G_{xx}(\omega)$ takes real values only on the imaginary axis, where it behaves monotonically from $\tilde G_{xx}(0)$ to $\tilde G_{xx}( i \infty)$ (see \cite{Landau}, chapter XII). As we will show, $G_{xx}(0)=0$ in the cases of study and at infinite frequency it diverges, therefore the only pole coming from the zeroes of $\tilde G_{xx}$ is at the origin and can be avoided by choosing an appropriate integration contour in the complex plane.

Now, following \cite{drewcoleman}, consider the Kramers Kronig relation\footnote{Since $t_H=G_{xy}/G_{xx}$ and in the large frequency limit $G_{xy}\to0$ and $G_{xx}$ diverges, then $t_H^\infty=0$.}
\be
\omega\, \im [t_H(\omega)] = \frac{1}{\pi} \cp \int_{-\infty}^\infty \frac{\omega\, \re [t_H(\mu)]}{\omega-\mu} \dd \mu \, .
\ee
Defining $\omega_H = \lim_{\omega\to\infty} \omega\, \im[ t_H(\omega)]$, in the large $\omega$ limit we obtain
\be\label{eq:hallfreqdef}
\omega_H =  \frac{1}{\pi} \cp \int_{-\infty}^\infty  \re [t_H(\mu)] \dd \mu \, .
\ee
This constant $\omega_H$ is interpreted as the Hall frequency in \cite{drewcoleman} and gives the rate of the precession of charged particles in the magnetic field (note that with the AC current the charged particles will only move around a small part of this circle during any given oscillation of the electric field). Thus the Hall angle together with the Hall frequency can tell us about the helical motion of a particle in the setup of interest as discussed in more detail below. As in the case of the plasma frequency, we use the sum rule to define the Hall frequency and calculate it in a holographic setup; we are not demonstrating that both quantities are the same. This would have to be checked with an independent calculation. It is also possible to apply the hydrodynamic sum rule to this quantity, obtaining
\be
\re[t_H(0)]= \frac{\sigma_{xy}}{\sigma_{xx}}\Bigg|_{\omega=0} = - \frac{1}{\pi} \cp \int_{-\infty}^\infty \frac{\dd \omega}{\omega} \im[t_H(\omega)] \, ,
\ee
where the DC ohmic and Hall conductivities can be calculated analytically in most cases.

We can understand the definition for the Hall angle by relating it to an injected current in the medium along, say, the direction $x$
\be\label{eq:currratio}
t_H(\omega) = \frac{\tilde G_{xy}(\omega)}{\tilde G_{xx}(\omega)} = \frac{\sigma_{xy}(\omega)}{\sigma_{xx} (\omega)} \frac{E_x}{E_x} = \frac{j_y(\omega)}{j_x(\omega)}\, ,
\ee
where $j_y$ is the induced Hall current along the direction $y$. Fourier transforming back into time domain we find
\be\label{eq:indcurr}
j_y(t) = \int_{-\infty}^\infty t_H(t-t') j_x(t') \dd t' \, ,
\ee
with $t_H(t)$ the Fourier transformed Hall angle. Therefore, the presence of a magnetic field forces the electric current to precess in a dynamic process described by the Hall angle. The hall frequency $\omega_H$ determines the value of $\lim_{t\to t'}t_H(t-t')$, hence describing this precession at very short times, and for a perfect fluid, where diffusive centers are absent, should be equivalent to the cyclotron frequency \cite{drewcoleman}. In the case we study in section \ref{sec:hall} this is not the case though, since we consider a probe approximation and the fixed background acts as an infinitely massive diffusive center.

As mentioned before, $t_H$ is analytic in the upper half-plane. The integral will converge in the case we study later since the real part of $t_H$ decays faster than $1/\omega$ at large frequencies\footnote{For this, it is crucial that we do \emph{not} subtract the large frequency behaviour of the optical conductivity.}.  However, this will not be true for the imaginary part, given the finiteness of the Hall angle from figure \ref{fig:hallfreq}   and the definition of the Hall frequency $\omega_H = \lim_{\omega\to\infty} \omega \,\im[t_H(\omega)]$.

\section{Conductivity sum rule for $Dp$/$D(p+4)$ intersections} \label{sec:effects}

The metric and dilaton corresponding to the decoupling limit of a stack of $N_c\gg1$ black $Dp$-branes is given in the string frame by \cite{Johnson:2003gi}
\be\label{eq:background}
\dd s^2 = H^{-1/2} \left( - f \dd t^2 + \dd\vec x_p^2 \right) + H^{1/2} \left( \frac{\dd r^2}{f} + r^2 \dd \Omega_{8-p}^2 \right)\, , \quad e^\phi = H^\frac{3-p}{4} \, ,
\ee
where $f=1-(r_h/r)^{7-p}$ and $H=(L/r)^{7-p}$. The additional RR form does not play a r\^ole in this paper. The matter of the theory transforms in the adjoint representation. By considering an intersection of the stack of $N_c$ $Dp$-branes with a stack of $N_f\ll N_c$ probe $Dq$-branes matter transforming in the fundamental representation can be added \cite{hep-th/0205236}. It was shown that, in order to preserve supersymmetry in the extremal case, $r_h\to0$, the dimensionality of the fundamental branes, $q$, has to be given by $q=\{p,\, p+2,\, p+4\}$. In the latter case the probe branes wrap the worldvolume directions of the $Dp$-branes plus an $S^3$ contained in the $S^{8-p}$ appearing in \eqref{eq:background} whilst in the other cases the intersection must contain a defect in the worldvolume directions of the $Dp$-branes. Working in a dimensionless radial coordinate $u=r_h/r$ the intersection in the latter case is summarized in the following table:
\begin{center}
\begin{tabular}{l|ccccc}
 {} & $t$ & $\mathbf{R}^p$  & $u$ & $S^3$ & $X_{5-p}$ \\
 \hline
 $Dp$ & $\times$ & $\times$ &  &  \\
 $D(p+4)$ & $\times$  & $\times$ &  $\times$  & $\times$  &  
 \end{tabular}
 \end{center}
 with the dynamics governed by the DBI action
 \be\label{eq:dbiaction}
 S_{DBI} = - N_f T_{D(p+4)} \int_{{\mathcal{M}}_{p+4}} e^{-\phi} \sqrt{-\det [g+2\pi \ell_s^2 F]}   \, ,
 \ee
 where $T_{D(p+4)}= 1/\left( (2\pi \ell_s)^{p+4} g_s \ell_s  \right)$ is the tension of the brane, $g$ the pull-back of the $10$-dimensional metric to the worldvolume of the probe brane ${\mathcal{M}}_{p+4}$, $\ell_s$ the string length and $F$ the field strength of the gauge field living on the probe brane.
 
The embedding profile of the probe branes can in general be specified by a single scalar function $\psi(u)$ which is related to one of the angles in the compact $X_{5-p}$ manifold (there are situations where the embedding must be written in terms of multiple scalar fields but we here specialize to the more generic cases). This can be specified by decomposing the metric of the $(8-p)$-sphere as
\be
\dd \Omega_{8-p}^2 = \frac{\dd \psi^2}{1-\psi^2} + \psi^2 \dd \Omega_{4-p}^2 + (1-\psi^2) \dd \Omega_3^2 \, .
\ee
In this case, the asymptotic behavior of the embedding function near the boundary, $u\to0$, is given by $\psi \approx m_q \,u + c_q \, u^3$. The dimensionless constant $m_q$ is directly related to the quark mass (see \eqref{eq:dimensmass}) and $c_q$ to the quark bilinear condensate \cite{hep-th/0701132}.

The system presents two different phases, separated by a first order phase transition at a critical mass $m_q^*$ \cite{hep-th/0701132}. For masses below the critical value a horizon is induced in the worldvolume of the flavor branes, giving rise to a conducting phase in which the mesons become unstable. For masses larger than the critical value the transverse $S^3$ wrapped by the branes shrinks to zero size above the horizon of the background geometry, and the system is in an insulator phase, with  a stable mesonic bound-state spectrum.

It is possible to introduce a finite baryon density into the description of the field theory by turning on the temporal component of the $U(1)$ abelian center living on the worldvolume of the probe branes. On the horizon this component has to satisfy $A_0(1)=0$, in order to avoid a singular one-form. From the asymptotic behavior on the UV, $ A_0 \approx \mu - a u^2$, we identify $\mu$ as the chemical potential and $a$ as related to the baryon density. Whenever $a\neq 0$, the embedding has to be of the black hole kind, but a phase transition still remains if the value of the baryon density is low enough \cite{hep-th/0611099}.

In the present work we are interested in the fluctuations of the $U(1)$ vector field in the conducting phase. The expression \eqref{eq:sumrule} is evaluated at null spatial momentum for the fluctuations, so an $SO(p)$ symmetry is preserved and the $p$ spatial components of the vector field can be studied by considering just one of them, say $\delta A_p(t,u) = \exp(-i \omega t) \delta \alpha_p(u)$, with $\delta A$ the perturbation of the gauge field. We will work in the $\delta \alpha_u=0$ gauge. The equation of motion for the fluctuation is obtained by expanding \eqref{eq:dbiaction} to quadratic order, and is integrated imposing that it is an ingoing-wave at the horizon of the black hole and fixing an overall normalization. In the UV it behaves as $\alpha_p(u\sim0)\approx {\cal A}_p + {\cal B}_p\, u^2 + \cdots$. An additional logarithmic term can appear in this expansion. We will see later how to deal with these.

One can now obtain the two-point function $\tilde G^{J^p,J^p}(\omega,0)$ using the standard prescription \cite{hep-th/0205051}  (from now on we will focus on the $J^p$-$J^p$ correlator, so we will skip this superscript)
\be\label{eq:prescription}
\tilde G(\omega) = 2 V(u) \frac{{\cal B}_p}{{\cal A}_p} \Bigg|_{u\to\epsilon}\, ,
\ee
where $V(u)$ is proportional to the wronskian for $\alpha_p$, and can be read from the $(\partial_u \delta\alpha_p)^2$ term in the quadratic action. Notice that it captures several constants like $N_f T_{D_{(p+4)}}$,  $(2\pi\ell_s^2)^2$ or the volume of the $S^3$.

\subsection{The $D3$/$D7$ system}

We first specialize to the $D3$/$D7$ brane intersection, corresponding to quenched fundamental matter added to  four dimensional (thermal) $\NN=4$ SYM with a small number of $\NN=2$ hypermultiplets.

In order to evaluate \eqref{eq:sumrule} we must subtract the $T=0$ contribution from the `bare' Green's function. We subtract the analytic part of the correlator at zero mass and zero baryon density which can be found in \cite{Myers:2007we}. Evaluating the retarded correlator at large real frequencies we obtain
\be\label{eq:d3d7asymp}
\tilde G_{S}(\omega) =  -  \frac{N_f N_c T^2}{4} \left(\frac{\omega}{2\pi\,T}\right)^2 \left( 2\gamma_E +\log\left(\varepsilon^2\, \left(\frac{|\omega|}{2\pi\,T}\right)^2\right)- i\, \pi\, {\mathrm{sign}}(\omega)\right) \, .
\ee
The $\varepsilon\to0$ limit should also be understood in the previous expression. The logarithmic divergence is avoided by the use of holographic renormalization. To make this more concise we can Fourier transform  \eqref{eq:d3d7asymp} back into the time-domain, obtaining a correlator
\be
G_{S}(t) = \frac{N_f N_c}{4\,\pi^2} \left(   \frac{1}{2} \log(\varepsilon^2) \ddot \delta(t) + \pi \partial_t^2 \frac{\Theta(t)}{t}\right) \, ,
\ee
and we see that the $T=0$ contribution is a retarded correlator with a contact term, which presents the logarithmic divergence. The $\ddot \delta(t)$ contact term could alternatively have been removed by adding the appropriate counterterm to the DBI action via holographic renormalization given by \cite{Karch:2007pd}
\be
L_F = -\frac{L^4}{4}N_f T_{D_7} V(S^3)  (2\pi\ell_s^2)^2 \int_{{\cal M}_4} \sqrt{-h}\, h^{il}h^{jm}F_{ij}F_{lm} \log(\epsilon^2) \Big|_{u\to \epsilon} \, ,
\ee 
where $h_{\mu\nu}$ is the $4$-dimensional metric restricted to the $u=\epsilon$ slice. The subtraction of \eqref{eq:d3d7asymp} is equivalent to the addition of this counterterm. However, the term proportional to $\partial_t^2 \Theta(t)/t$ cannot be removed with holographic renormalization and is needed to regularize the spectral function at large frequencies.
 
\begin{figure}
\begin{center}
\epsfig{file=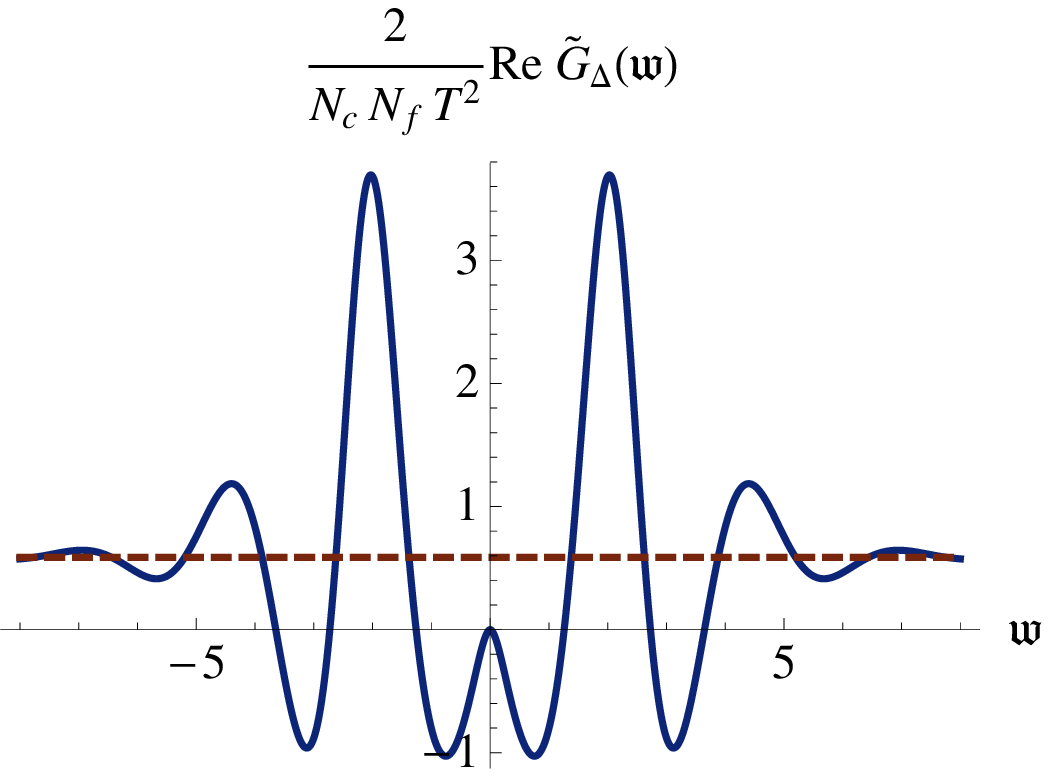,scale=0.65}
\epsfig{file=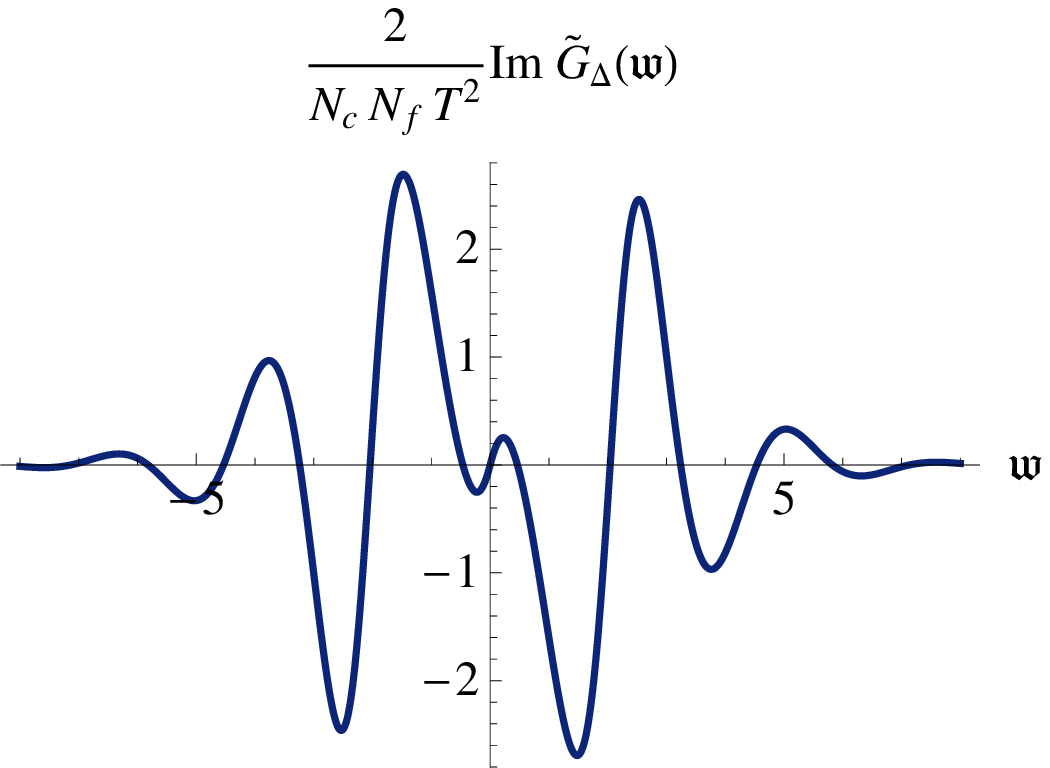,scale=0.65}
\caption{\em \label{fig:wcorrelator}
Real (left) and imaginary (right) parts of the correlator function for finite mass $m_q=1.3266$, finite baryon density $\Dis=2$ and zero momentum, in units of $N_c N_f T^2/2$. The dashed line marks the value $m_q^2/3\approx0.58662$.
}
\end{center}
\end{figure}

Having tamed the divergences, we can investigate the finite temperature correlator as a function of the frequency in a representative range of masses and baryon densities\footnote{The dimensionless baryon density $\Dis$ that we specify in the examples is related to the $a$ constant in the asymptotic expression of the $A_0$ mode by $a=\frac{\pi T L^2}{2\pi \ell_s^2} \Dis$.}. We present a typical profile for the retarded correlator in figure \ref{fig:wcorrelator}, and we will refer to this figure as representative of the whole set of correlators we have studied.

One first thing to notice in figure \ref{fig:wcorrelator}(left) is that $\re\, \tilde G_\Delta(0)=0$. This is a result that has been obtained analytically in \cite{arXiv:0811.1750} to be $\tilde G(\omega) = i \omega \sigma_{DC}+ {\mathcal{O}}(\omega^2)$,  just reflecting the consistency of the numerical calculation of the retarded Green's function with the hydrodynamic limit. In the former expression $\sigma_{DC}$ is the conductivity of the system, which coincides with the macroscopic calculation in \cite{Karch:2007pd}.

We need now to calculate the value of the retarded correlator in the infinite frequency limit and check that it coincides with the r.h.s. of \eqref{eq:sumrule}. Studying this for different values of the quark mass and baryon density we find that the results are consistent\footnote{We found differences between the numerical and analytic values of only $\co(10^{-3}-10^{-4})$.}, and that
\be\label{eq:d3d7rhs}
 \tilde G^\infty_\Delta  =  \re\, \tilde G_\Delta(0,0) + \frac{1}{\pi} \cp \int_{-\infty}^\infty \frac{\im\, \tilde G_\Delta(\omega,0)}{\omega}\dd \omega\,=\frac{N_f N_c T^2}{6} m_q^2 = \frac{2N_f N_c}{3 \lambda} M_q^2\, ,
\ee
with $M_q = \frac{1}{2} \sqrt{\lambda} T m_q$, the dimensionful mass \cite{hep-th/0701132}. This result, up to a factor $2$, is the same as that obtained in appendix B of \cite{Myers:2007we} for the $T=0$ scalar-scalar correlator. In this limit both the scalar and vector channels are related by supersymmetry.

One could argue now that our subtraction scheme is incomplete, and that we should have also included the $M_q^2$ part in \eqref{eq:d3d7asymp}, which corresponds to a contact term proportional to $\delta(t)$. This would have changed the value of $\re\, \tilde G_\Delta(0)$ to a negative value, a fact with two important consequences. The first one is that the result would have been inconsequent with the hydrodynamic calculation, since this calculation is insensitive to the existence of such a contact term. The second consequence is however more crucial. A negative value of the correlator at zero frequency reflects, via the relation with the conductivity, in a $1/\omega$ pole in the imaginary part of the conductivity. The Kramers Kronig relations tell us that we should therefore find a delta function at zero frequency in the real part of the conductivity. This situation, however, cannot be realized within the probe approximation, since the delta function signals a saturation in the energy absorption of the system, but in the strict probe approximation limit the background is always able to absorb the energy pumped by the fields on the probe branes.

Coming back to \eqref{eq:d3d7rhs}, in the $m_q\to0$ limit we recover the R-current result\footnote{In this limit the R-current results are obtained by the rescaling $N_fN_c\to N_c^2$.} given by equation \eqref{eq:baier}. Also, this expression holds independent of the quark condensate or the baryon density. In particular, this means that the relation does not feel the unstable phases present in the vicinity of the phase transition, and will still hold if we overheat the system beyond the critical point, provided we do not reach the temperature at which the first quasinormal mode becomes tachyonic  \cite{hep-th/0701132}, since then the dispersion relations will not hold due to the presence of poles in the upper half plane.

We have also checked that the relations \eqref{eq:disprel1} and \eqref{eq:disprel2} hold, obtaining numerically figure \ref{fig:wcorrelator}(left) from \ref{fig:wcorrelator}(right)  and vice versa.  This can be used as a check of the numerical calculation, since it means that the signals in the system propagate causally, a fact that numeric instabilities in the integration of the equation of motion could spoil.

\begin{figure}
\begin{center}
\epsfig{file= 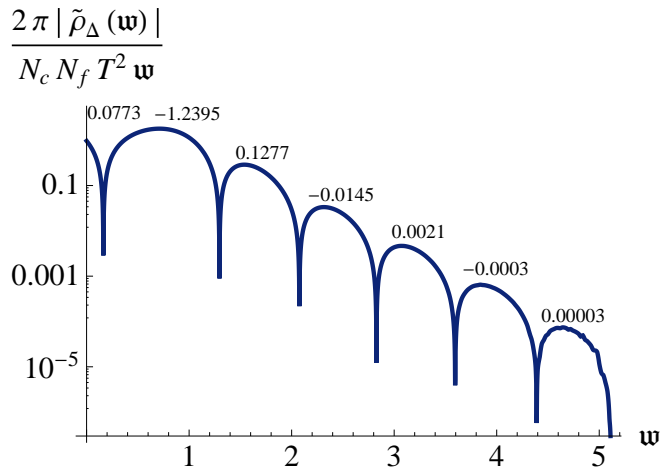,scale=0.82}
\caption{\em \label{fig:logintegrand}
Logarithmic plot of the integrand for the sum rule (in units of $N_f N_c T^2/(2\pi)$) as a function of ${\textswab{w}}=\omega/(2\pi T)$ for $\mass=1$ and $\Dis=0.1$. The amplitude of the oscillations decays as $\omega^2\! e^{-k\, \omega}$. The numbers above the line are the contribution to the integral by each oscillation, adding up to approximately $-\pi/3$, in agreement with the dispersion relation.
}
\end{center}
\end{figure}

An example of how numerical instabilities do not affect our results to any great extent is given by figure \ref{fig:logintegrand}. In this figure we show the absolute value of the integrand in \eqref{eq:disprel1} for a given value of the quark mass and baryon density, together with the value of the integration in the positive frequency axis. Comparing the value of this integral with the value given by \eqref{eq:d3d7rhs}  we obtain an overwhelming agreement, meaning that although we are not numerically integrating to infinite frequencies, the vanishing error introduced does not affect our results or conclusions. This is because the amplitude of the oscillations around the subtracted $T=0$ contribution decays exponentially, as observed in \cite{Teaney:2006nc,CaronHuot:2009ns}

In figure \ref{fig:logintegrand} it is seen that the last plotted bump already presents some small numerical errors. This error can be decreased with improved numerical methods.

\subsubsection*{Integrated conductivity}

The results described above can be expressed in terms of the sum rule for the conductivity \eqref{eq:specweight} by stating that
\be
W_{m_q} = \lim_{\omega_\Lambda\to\infty}W(\omega_\Lambda) =  - \frac{N_f N_c T^2}{6} m_q^2\, ,
\ee
where we have specified the value of the mass of the fundamental matter. Recalling the sum rule \eqref{eq:plasmon} we observe that the effect of introducing massive fundamental matter in the system is to effectively \emph{reduce} the plasma frequency. The more massive the fundamental matter, the bigger reduction we get, though clearly this is a subleading effect in $N_f/N_c $.

\subsection{$Dp$/$D(p+4)$ systems}

We will restrict the discussion in the current section to $p=\{1,\, 2,\, 3, \, 4\}$, since the background created by $Dp$-branes with $p\geq5$ is thermodynamically unstable \cite{Cai:1999xg} and we cannot define the current-current correlator at zero momentum, equation \eqref{eq:prescription}, for $p<1$. These setups are the holographic duals to $p+1$-dimensional theories, and therefore its interest is mostly academic in the $D4$/$D8$ case (note that we are not introducing compact directions for the $Dp$-brane world volume directions). $p=1$ and 2 may be used to model some condensed matter systems.

We have not found in the literature the equivalent to the asymptotic expression \eqref{eq:d3d7asymp} for $p=\{2,\, 4\}$, which is needed to define the subtracted Green's function $\tilde G_\Delta(\omega)$ entering in the expression for the sum rules. When $p=1$ the integration of the vector mode can be performed analytically in the massless, zero baryon density case, obtaining a solution satisfying the ingoing-wave condition at the horizon and normalized at the boundary
\be\label{eq:d1d5corr}
\alpha_1 (u) = \exp\left[  -i \frac{3\, \omega}{2\pi\,T} \int_0^u \frac{\hat u\, \dd \hat u}{1-\hat u^6} \right] \, .
\ee
From here it is straightforward to see that the retarded correlator is $\tilde G(\omega) \propto -i \frac{3\, \omega}{2\pi\,T}$. This is precisely the quantity we should subtract, since it diverges at large frequencies and corresponds to nothing but a $\dot\delta(t)$ contact term. It is then clear that, in this case, the sum rule in the massless, zero baryon density case is zero, since the integrand vanishes identically. We found the result for the massless sum rule to be independent of the baryon density.

We do not have an equivalent expression to \eqref{eq:d3d7asymp} or \eqref{eq:d1d5corr} for the $D2$/$D6$ and $D4$/$D8$ cases. Despite of this, we can evaluate the conductivity sum rule by using experience learnt from the $D1$/$D5$ and $D3$/$D7$ systems. This we can do by changing to an alternative subtraction scheme, in which we do not only subtract the analytic divergent part of the correlator at $T=0$, but take away the whole Green's function corresponding to the massless, zero baryon density case at \emph{finite} temperature. With this, we are adding some frequency-dependent structure to the spectral density of $\tilde G_\Delta$, originating from the introduction of an infinite set of quasinormal modes (in the complex plane).

The new subtraction scheme redefines $\tilde G_{\Delta}(\omega) = \tilde G(\omega, m _q, \tilde d, T)- \tilde G(\omega, 0,0,T)$, as opposed to the definition given in section \ref{sec:disprels}, which assumed also $T=0$ in the subtracted piece. Therefore, the conductivity sum rule \eqref{eq:specweight} evaluates now to $W_{m_q,\tilde d} = -\tilde G^\infty_{\Delta}(m_q, \tilde d) + \tilde G^\infty_{\Delta}(0,0)$. However, if the aim is to  compare the integrated conductivity for two different masses, this shift will have no effect and our results are not affected by choosing this subtraction scheme. In the $D3$/$D7$ and $D1$/$D5$ cases we found that $G^\infty_{\Delta}(0,0)=0$, and we will consider that this holds in the following. This seems a reasonable assumption, since the system has not a natural scale to provide the dimensions for $\tilde G^\infty_{\Delta}(0,0)$.

Performing a scan of the mass and baryon density parameter space, we find that the large frequency behavior of the integrand in \eqref{eq:specweight} again decays exponentially in the $D4$/$D8$ case, whereas in the $D1$/$D5$ and $D2$/$D6$ systems it decays just with a power of the frequency ($\omega^{-2}$ and $\omega^{-7/3}$ respectively). However, these systems are more stable numerically, and we can obtain reliable results at large frequencies. The calculation of the spectral sum rule can be summarized by the expression
\be\label{eq:specweightforp}
W_{m_q} =  - 2\pi T\, \sigma_{0} \,\kappa_p\, m_q^2\, ,
\ee
where $\sigma_{0}= N_f N_c \, \NN_p $ is the DC conductivity at zero mass and zero baryon density,
and the values for $\NN_p$ and $\kappa_p$ are given by
\begin{center}
\begin{tabular}{l||c|c}
$p$ & $\kappa_p$ & $\NN_p$  \\
\hline
$1$ & $0.598(3)$ & $ 1/\sqrt{\pi\, \lambda}$ \\
$2$ & $0.713(2)$ & $ \left(9\, T/160 \pi\lambda\right)^{1/3}$ \\
$3$ & $0.333(1)$ & $T/4\pi $ \\
$4$ & $1.199(1)$ & $\lambda\,T^3/54\pi$ 
\end{tabular}
\end{center}

$\kappa_p$ is the parameter we have determined numerically.  Notice that the conductivity  has energy dimension $\delta_\sigma=p-2$, whereas the 't Hooft coupling, $\lambda=N_c \,g_{YM}^2$, has energy dimension $\delta_\lambda=3-p$ in each case. These expressions can be written in terms of the dimensionful mass \cite{hep-th/0701132}
\be\label{eq:dimensmass}
M_q = \frac{r_h}{2^{\frac{9-p}{7-p}}\pi\ell_s^2}m_q\, ,
\ee
such that
\be
W_{M_q} = -\beta_p N_f N_c \frac{M_q^2}{\lambda} \, ,
\ee
where $\beta_1=3\kappa_1/2^{1/3}$, $\beta_2=5\kappa_2/2^{6/5}$, $\beta_3=2\kappa_3$ and $\beta_4=3\kappa_4/2^{2/3}$.

\section{Conductivity sum rule in intersections with a defect}\label{sec:defects}

Up to now we have studied theories where the intersecting dimensions of the $Dp$/$Dq$ system were precisely those of the $Dp$-brane worldvolume, however, it is possible to study similar setups, where a defect does exist in the intersection. One of the possibilities is the Sakai-Sugimoto model \cite{Sakai:2004cn}, which has received lots of attention as a holographic model of QCD. However, it is not clear how a finite quark mass would be added, so we cannot use this system to find a similar result to \eqref{eq:specweightforp}. Evaluating the sum rule in the massless case we obtain that the integrated conductivity  vanishes.

Another candidate system to describe properties of $3+1$ plasmas is a $D4$/$D6$ intersection with one compactified dimension \cite{hep-th/0701132}. This intersection is summarized in the following table
\begin{center}
\begin{tabular}{l|cccccc}
 {} & $t$ & $\mathbf{R}^3$ & $x^4$ & $u$ & $S^2$ & $X_2$ \\
 \hline
 $D4$ & $\times$ & $\times$ & $\times$  &  &  \\
 $D6$ & $\times$  & $\times$ &   & $\times$  & $\times$  &  
 \end{tabular}
\end{center}
As in the $D3$/$D7$ case, it is possible to add a finite baryon density and perturb the gauge field to obtain the retarded current-current correlator. We have performed a study calculating the sum rule for different values of the mass and the baryon density and found that the integrated conductivity vanished always. This suggest that the presence of a defect dimension in the theory has a very important effect on the determination of the sum rules.

We studied also the case of a $2+1$-dimensional theory with one defect, namely the $D3$/$D5$ intersection with a non trivial flux along the defect to stabilize the setup  \cite{Myers:2008me}. The intersection is
\begin{center}
\begin{tabular}{l|cccccc}
 {} & $t$ & $\mathbf{R}^2$ & $x^3$ & $u$ & $S^2$ & $X_3$ \\
 \hline
 $D3$ & $\times$ & $\times$ & $\times$  &  &  \\
 $D5$ & $\times$  & $\times$ &   & $\times$  & $\times$  &  
 \end{tabular}
 \end{center}
Considering again the current-current correlator and scanning the parameter space of the theory for several values of the mass and baryon density we find that the integrated conductivity always vanished in contrast to the non-defect theory.

Hence, we find that in flavor models based on intersections at a defect dimension there is no shift in the plasma frequency due to the presence of fundamental degrees of freedom.

\section{The frequency dependent hall conductivity and associated sum rules}\label{sec:hall}

In this section we calculate the frequency dependent Hall conductivity and associated sum rules in the case of the $D3$/$D7$-brane intersection\footnote{See \cite{Alanen:2009cn} for a similar calculation in a setup with a defect.}. In order to find the Hall conductivity in a macroscopic setup one must turn on both an electric and perpendicular magnetic field and find the current induced in the transverse direction to these external fields. Using Ohm's law one may then deduce the conductivity. However, such a setup will in general only provide the DC hall conductivity. This has been done previously in \cite{O'Bannon:2007in}. Here we are interested in the frequency dependent (AC) conductivity, for which a microscopic calculation is needed. The setup is therefore as follows. A probe $D7$-brane is embedded in a black $D3$-brane background and a gauge field of the following form is turned on
\begin{equation}
A=\frac{A_t(u)}{2\pi\alpha'} \dd t + e^{-i \omega t}\delta \alpha_x(u) \dd x+  \left( e^{-i \omega t} \delta \alpha_y(u)+\frac {\pi^2 T^2 L^2}{2\pi\alpha'} \tilde B \,x\right) \dd y \, .
\end{equation}

The gauge field therefore encodes a magnetic field in the $z$-direction, a baryon density given by the temporal component of the gauge field and excitations of the gauge field in the $x$ and $y$ directions. Note that the choice to include only zero momentum modes will simplify the calculation a great deal as it allows us to decouple scalar and vector fluctuations.

The combination of the magnetic field and the finite baryon density causes a coupling between the $x$ and $y$ components of $\delta \alpha_\mu$, and thus solving the differential equations becomes much more complicated. However, it is possible to chose an appropriate basis  for the gauge field in which we can decouple the equations of motion. This is the basis of circular polarizations which we label $\alpha_L(u)$ and $\alpha_R(u)$, standing for left and right circular polarizations. This change of basis is implemented by
\begin{equation}
\delta \alpha_x(u)=\frac{\delta \alpha_L(u)+\delta \alpha_R(u)}{2}\, , \quad \delta \alpha_y(u)=\frac{\delta \alpha_L(u)-\delta \alpha_R(u)}{2i} \, .
\end{equation}

From the usual DBI action, we Legendre transform with respect to the time component of the gauge field at zeroth order in the fluctuations in order to calculate the embedding profile for the $D7$-brane. Such a calculation has been shown in detail numerous times (see for example \cite{hep-th/0611099}). In all of what follows we will be at finite baryon density, as the Hall conductivity is only present in this case. In the presence of finite baryon density all embeddings are of black hole type and end with the $D7$-brane on the black hole horizon. 

In order to calculate the fluctuation solutions for the $L$ and $R$ components of the gauge field, we expand the action to quadratic order in these fields allowing a plane wave ansatz for the solution. Off-shell there are terms in the action which are quadratic in both the $\delta \alpha_L$ and $\delta \alpha_R$ fields as well as cross terms which mix the two. However at the level of the equations of motion, where the embedding solution is put on-shell, this coupling in the equations vanishes ({\it{i.e}}. the coupling terms in the fluctuation e.o.m. vanish if we plug the solution of the embedding profile). In fact the coupling at the level of the action becomes pure and the $L$ component equation of motion involves only the $R$ fluctuation and vice versa.

Having uncoupled the equations of motion we can calculate the Green's function with a numerical integration, by specifying the incoming wave boundary conditions at the horizon and solving out to the UV where we read off the asymptotic behaviour. However, although the fluctuations decouple in the equations of motion, they couple exactly (there are no non-coupled terms) in the on-shell 
 action. The Green's function, as a matrix in the space of left and right circular polarizations, is therefore exactly off-diagonal. The UV behaviour of the solutions is of the form
\begin{equation}
\delta \alpha_{L,R}(u)|_{u\rightarrow 0}=a_{L,R}+b_{L,R}u^2 \, .
\end{equation}
By studying the UV behaviour of the action therefore one can determine the exact form of the Green's function matrix, which is given by
\begin{equation}
\begin{pmatrix}
\tilde G_{LL}    & \tilde G_{LR} \\
\tilde   G_{RL}  & \tilde G_{RR} 
\end{pmatrix}
\sim
\begin{pmatrix}
 0    &   \frac{b_L}{a_L}\\
   \frac{b_R}{a_R}  &   0
\end{pmatrix} \, .
\end{equation}

We can study the matrix of Green's functions in the original $(x,y)$ basis  by transforming the matrix in the $(L,R)$ basis, which is calculated from the numerical calculation. The $(x,y)$ Green's function matrix is then given by
\begin{equation}\label{eq:Greensmat}
\begin{pmatrix}
\tilde G_{xx}    & \tilde G_{xy} \\
\tilde   G_{yx}  & \tilde G_{yy} 
\end{pmatrix}
=\frac{1}{4}
\begin{pmatrix}
\tilde G_{RL}+\tilde G_{LR}   &   i(\tilde G_{LR}- \tilde G_{RL})  \\
  -i(\tilde G_{LR}- \tilde G_{RL})    &   \tilde G_{RL}+ \tilde  G_{LR}
\end{pmatrix}\, .
\end{equation}
This defines the bare green's function, and so in order to study the Kramers Kronig relations we must subtract the large $\omega$ divergent behavior as discussed in the previous sections. Notice that, naturally, $\tilde G_{xx}=\tilde G_{yy}$ and $\tilde G_{xy} = -\tilde G_{yx}$.

The equations of motion for $\delta \alpha_R$ and $\delta \alpha_L$\footnote{Details of the equations of motion can be found in equation (A.2) in \cite{Tarrio:2009vk} .} are equal up to a sign change of $\tilde B$, as expected, and thus the above Green's function is governed by a single equation, where we must chose the appropriate sign of $\tilde B$ to perform the calculation. From the Green's function the conductivity follows simply and thus we are left with the final expression for the complex hall conductivity
\begin{equation}
\sigma_{xy}=\frac{1}{4\, \omega}\left(\frac{b_L}{a_L}-\frac{b_R}{a_R}\right)\, .
\end{equation}
Note that the large $\omega$ divergence cancels in this expression and thus the large $\omega$ subtraction is unnecessary. This we will find not to be true for $\sigma_{xx}$. We can write the Hall conductivity alternatively as
\begin{equation}
\sigma_{xy}=\left.\frac{1}{4u\,\omega}\left(\frac{\delta \alpha'_L(u,B,\tilde d)}{\delta \alpha_L(u,B,\tilde d)}-\frac{\delta \alpha'_L(u,-B,\tilde d)}{\delta \alpha_L(u,-B,\tilde d)}\right)\right|_{u\rightarrow 0} \, ,
\end{equation}
where the primes are derivatives with respect to $u$. There are some immediate points to note about the calculation and the preliminary results. As explained in previous sections, the Green's functions should satisfy a Kramers Kronig relation. This is both a good check of our definitions and of the numerics of our computation. We have performed this comparison and found perfect agreement between the numeric calculation of the real (imaginary) part of the correlator and the result obtained via Kramers Kronig relations from the imaginary (real) part.

The DC limit of the above expression gives
\be
\sigma_{xx} = \frac{N_f N_c T}{4\pi}\sqrt{\frac{(1-\psi(1)^2)^3}{1+\tilde B^2} +  \frac{\tilde d^2}{(1+\tilde B^2)^2}}\, , \quad \sigma_{xy} =  \frac{N_f N_c T}{4\pi} \frac{\tilde B \, \tilde d}{1+\tilde B^2}\, ,
\ee
which was previously obtained in \cite{Tarrio:2009vk} and matches the macroscopic calculation in \cite{O'Bannon:2007in}. We have found that there is again a perfect fit. It should be noted that the $\psi$ independence in the DC Hall conductivity, explicit in the analytic result, is mirrored in the numerical solution. This is another very good check that both the conventions for defining the Green's function and the numerics themselves are correct.

\begin{figure}
\begin{center}
\epsfig{file= 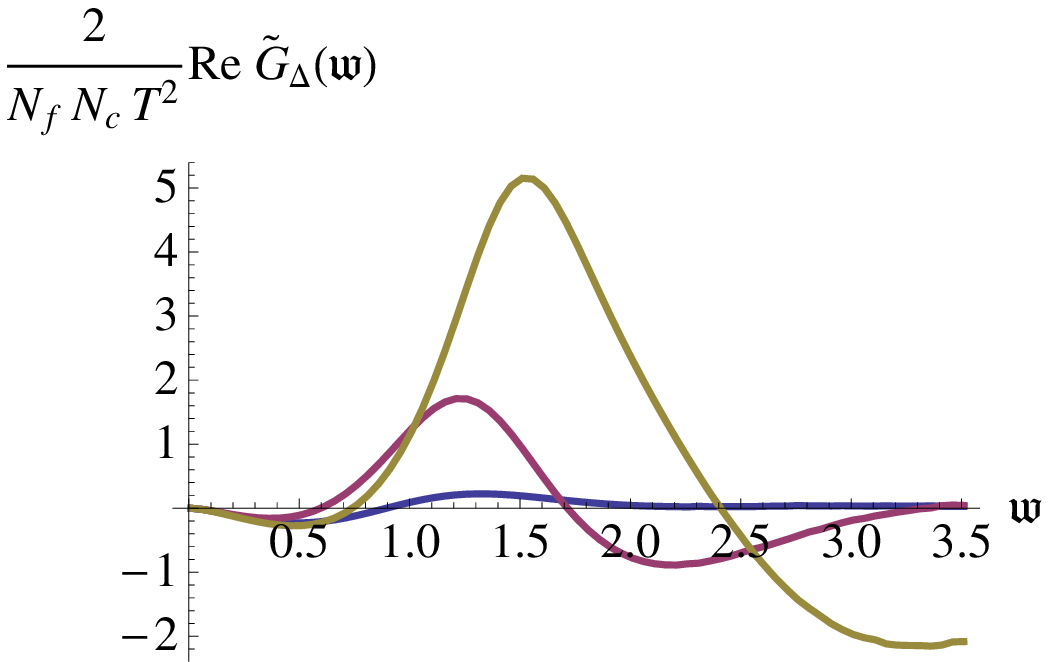,scale=0.65}
\epsfig{file= 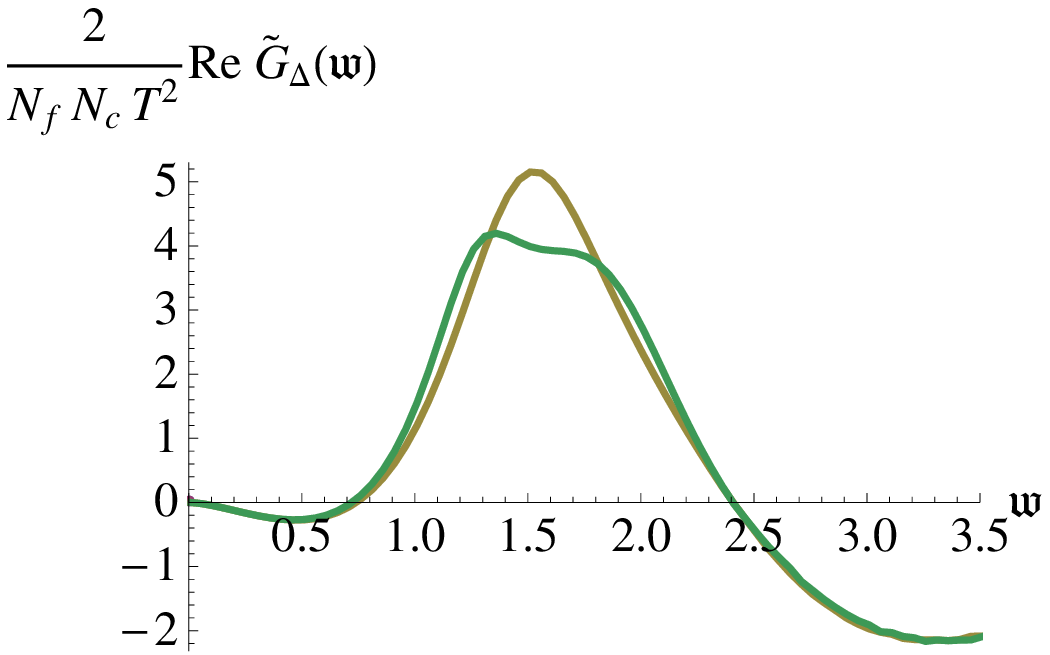,scale=0.65}
\caption{\em \label{fig:zeeman}
(Left) Real part of the regularized Green's function for fixed mass $m_q=0.297$, zero baryon density and increasing magnetic field $\tilde B=0$ (blue), $\tilde B=2$ (red) and $\tilde B=5$ (yellow). (Right) Real part of the regularized Green's function for fixed magnetic field $\tilde B=5$, zero baryon density and increasing angle of incidence of the embedding profile in the black hole, corresponding to masses $m_q=0.297$ (yellow) and $m_q=0.256$ (green). Note that the values of $\psi_0$ are 0.72 and 0.9 respectively. For the larger mass, smaller $\psi_0$, the Zeeman effect is hidden by the broader peaks in the spectral function.
}
\end{center}
\end{figure}

In figure \ref{fig:zeeman} we plot the real part of the $x-x$ Green's function for specific values of the magnetic field, baryon number density and quark mass as a function of the frequency. There is a clear feature appearing in these plots which is not present in the case of zero baryon density and/or zero magnetic field. This is the appearance of double peaks, which are clearly almost coincident, and the distance apart of which is related to the size of $\tilde B$ and $\tilde d$. We can study the exact position of these peaks by studying the poles in the spectral function in the complex frequency plane, and see clearly the appearance of two peaks, emerging  from a single peak for $\tilde B=0$ or $\tilde d=0$. This effect of the splitting of a single quasinormal mode into two is the finite temperature analogue of the Zeeman effect. The breaking of parity symmetry in our system by the magnetic field causes a splitting in parity pairs of the quasinormal spectrum. This splitting can be seen explicitly in \cite{Filev:2007gb, Filev:2010pm}.

Having studied the spectral function itself we can now look at `global' properties of the Green's functions, defined through sum rules. The first quantity to discuss is the Hall angle, given by the ratio of the Hall conductivity to the optical conductivity
\begin{equation}\label{eq:tH}
t_H(\omega)=\tan \theta_H(\omega)=\frac{\sigma_{xy}(\omega)}{\sigma_{xx}(\omega)} \, .
\end{equation}
\begin{figure}
\begin{center}
\epsfig{file= 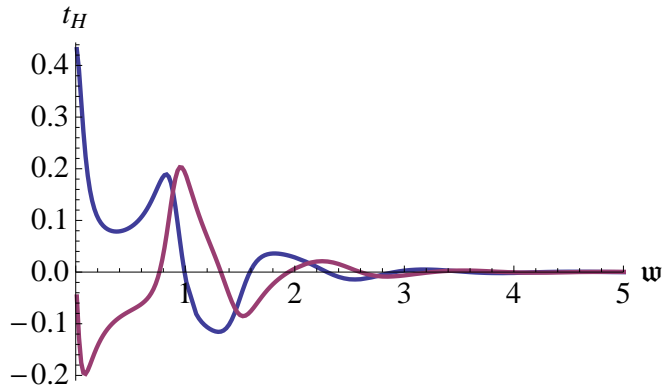,scale=0.82}
\caption{\em \label{fig:hallangle}
Real (blue) and imaginary (red) parts of the Hall angle $t_H$ for $m_q=1.06$, $\tilde B=0.5$, $\tilde d=0.1$.}
\end{center}
\end{figure}
An example of this quantity is plotted in figure \ref{fig:hallangle}. It has a number properties which differ greatly from one material to another. In particular the hall angle tells us about the ratio of Hall to drift currents under a given electric field  \eqref{eq:currratio}.

\begin{figure}
\begin{center}
\epsfig{file= 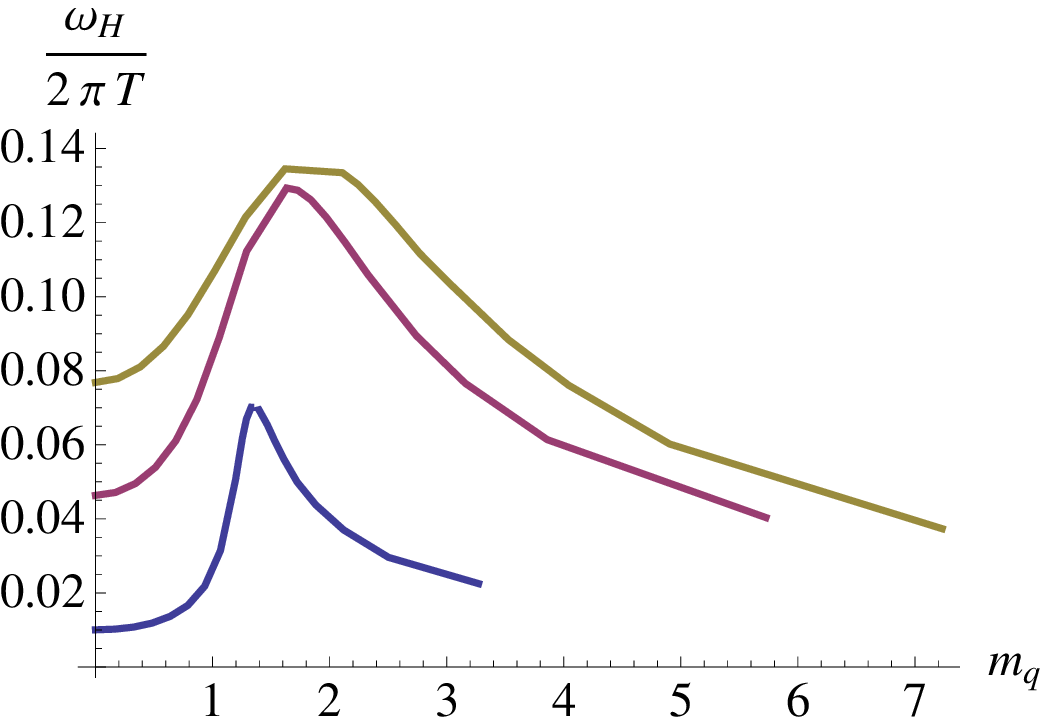,scale=0.65}
\epsfig{file= 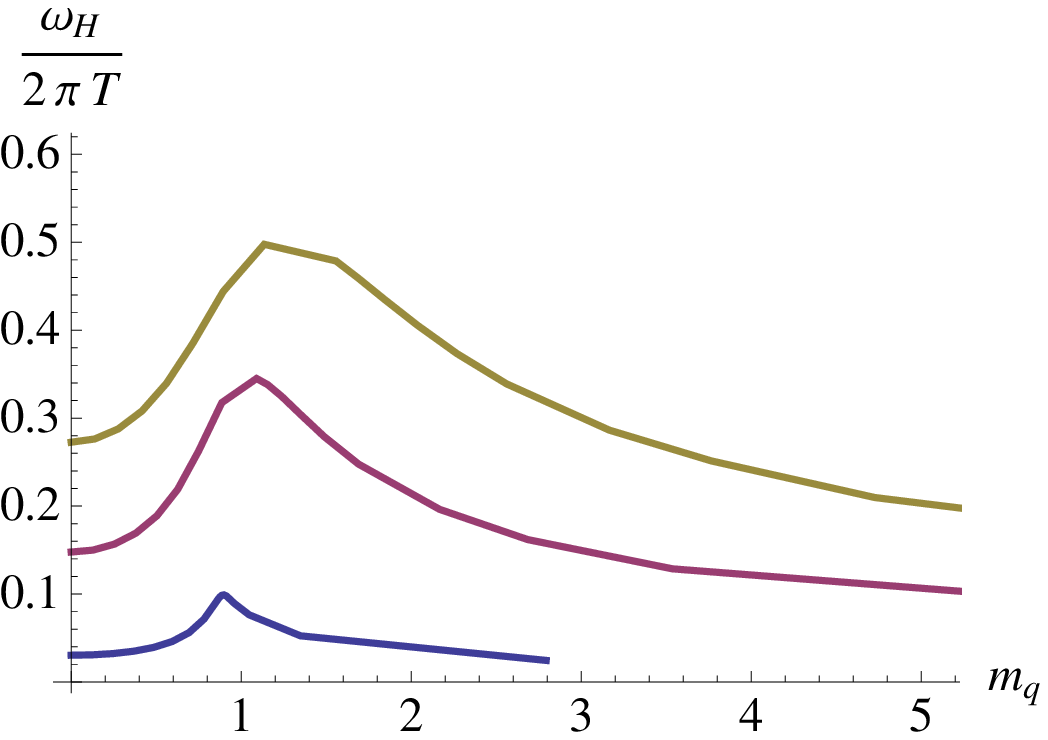,scale=0.65}
\epsfig{file= 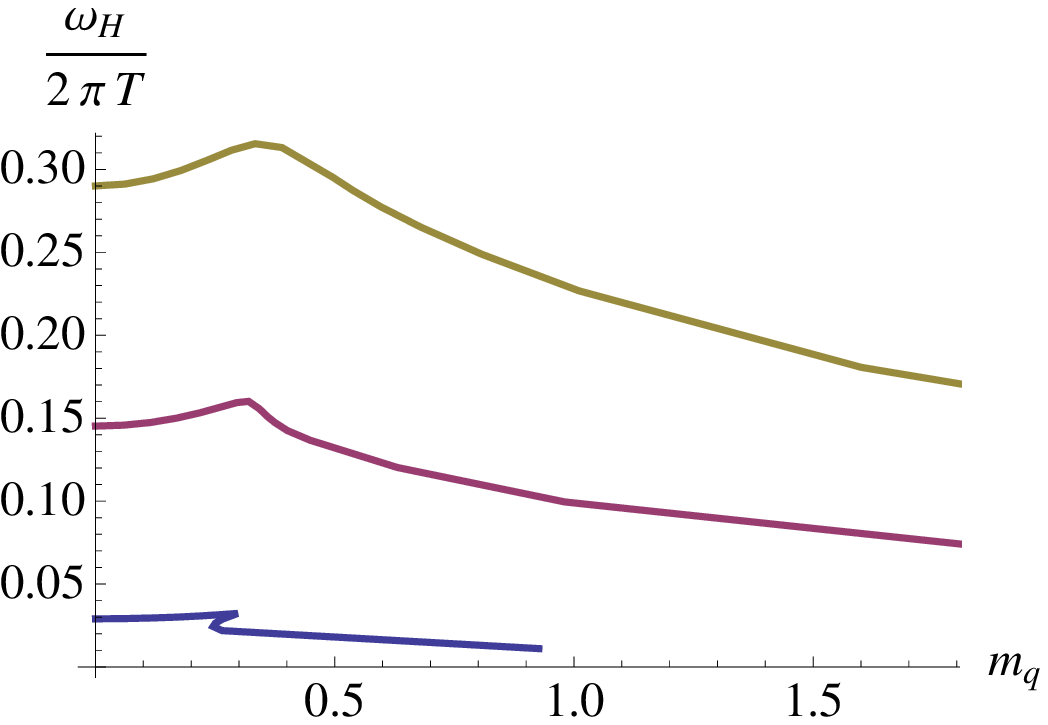,scale=0.65}
\epsfig{file= 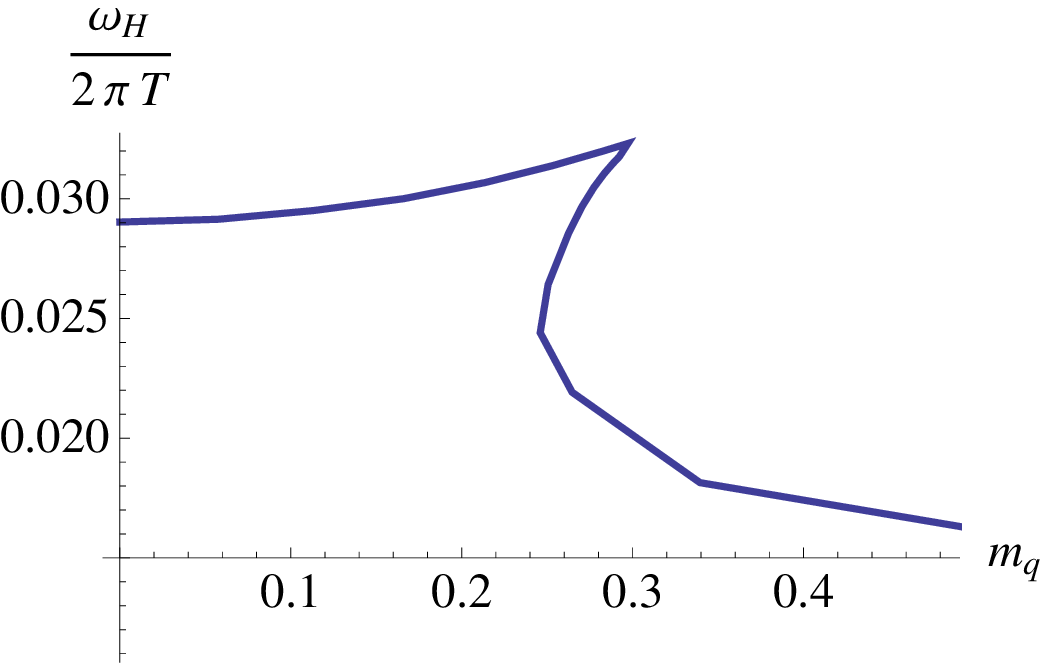,scale=0.65}
\caption{\em \label{fig:hallfreq}
Hall frequency obtained from the Hall angle sum rule as a function of the mass for $\tilde d=0.1$ (blue), $\tilde d=0.5$ (red) and $\tilde d=1$ (yellow) and $\tilde B=0.5$ (top-left figure),  $\tilde B=2$ (top-right figure),  $\tilde B=5$ (bottom-left figure). The Hall frequency tells about the relative strength of the optical to Hall conductivities in the large frequency domain and thus we see that there is a, for a fixed $\tilde B$ and $\tilde D$, a value of the mass for which the relative strengths are closest. Below and above this value the optical conductivity dominates the transport of charge carriers in the material. The bottom-right figure corresponds to the blue curve in the bottom-left one. We observe the multivaluedness signaling the presence of a phase transition for these values of the magnetic field and charge density.
}
\end{center}
\end{figure}

From the hall angle we can define the hall frequency, $\omega_H$, given by \eqref{eq:hallfreqdef}. In figure \ref{fig:hallfreq} we present the numeric results we obtained for the value of the Hall frequency as a function of the external magnetic field, the baryon density and the mass of the fundamental degrees of freedom. In the examples we present we can see that, for a given $\tilde B$ and $\tilde d$, the Hall frequency has a maximum value for an specific mass $m_q$. After this maximum is reached, the Hall frequency decays with an inverse power of the quark mass. We couldn't find any analytic expression for this power or the position of the maximum.

The strange feature observed in  figure \ref{fig:hallfreq}(d) is reminiscent of a phase transition \cite{Filev:2007gb,Erdmenger:2007bn,Evans:2010iy}. We can see that the Hall frequency is sensitive to this change of phase, which occurs for $m_q\approx 0.27$ for $\tilde B=5$ and $\tilde d=0.1$. It is not a surprise that the Hall frequency has a jump at the value of the critical mass, since the system goes to a chiral setup when the magnetic field is increased, and we expect the diffusive processes of the plasma to be modified.

\section{Conclusions and perspectives}\label{sec:conclu}

It would be interesting to find general sum rules for the retarded correlators in the context of AdS/CFT. The biggest problem is the evaluation of the asymptotic value $\tilde G^\infty_\Delta$. As pointed out in
\cite{CaronHuot:2009ns}, these values are related to expectation values of local operators via  OPEs in the field theory side. Also, it is expected that the results at weak and strong coupling differ considerably, as shown in \cite{Baier:2009zy} for $\NN=4$ SYM. Comparison with results from lattice in the study of thermal plasmas will certainly shed light on whether we should trust holographic estimates of transport coefficients from the spectral functions, although it is known that the calculation of real time thermal properties on the lattice is extremely difficult. In \cite{Gulotta:2010cu} the authors are able to find analytic expressions for a number of sum rules in a gravitational context, and it would be interesting to check whether the calculations presented here can be merged with their approach.

This tool may also be useful in the AdS/CMT approach, in which the determination of the properties of the AC conductivity is playing an important r\^ole. It may also be helpful for determining the strength of the delta functions appearing at zero frequency, and in this form give information about the gap frequency of the models under consideration.

In this paper we have calculated sum rules related to both the optical and hall conductivities for probe flavors in the plasma of adjoint matter. The results that we obtain tell us about how fundamental matter affects these transport properties in the quenched approximation. It is know that the probe limit, which has been assumed throughout, does not allow one to follow the dynamics of such processes for infinitely long times, as the effects of the fundamental matter on the adjoint degrees of freedom will eventually become non-trivial. In the preceding timescale however we are able to calculate finite conductivities and thus finite plasma frequencies and hall conductivities, both in the DC and AC regimes (indeed, as shown, the limit of our AC results agree with the previously calculated DC results which are found through classical D-brane solutions in the absence of fluctuations and do not rely on the regularization of Green's functions). The fact that the sum rules link the well established DC results, to the global properties of the AC results is a strong indication that these calculations are correct.

We have calculated both the plasma frequency and hall frequency which in such a setup tell us about the response of fundamental matter to microscopic AC electric currents in addition to finite magnetic fields and baryon densities. Although clearly such fine properties are not accessible to heavy ion experiments, the techniques outlined here allow for similar analysis in lower dimensional systems which should be comparable to condensed matter systems. The full analysis of these systems is in progress.

The fact that we are able to extract highly non-trivial properties of strongly coupled systems via holographic sum rules takes us another step closer to calculating measurable quantities in real-world systems. The backgrounds discussed in this paper are the simplest we could study, but the results here can be extended quite easily to more realistic systems which one may eventually be able to engineer in the laboratory.

\section*{Acknowledgments}
We would like to thank Jorge Casalderrey-Solana, Sean Hartnoll, Carlos Hoyos-Badajoz, Matthias Kaminski and Aninda Sinha for comments and discussions. J. T. would like to thank the Perimeter Institute for hospitality when this project was started. J.S. and J.M. would like to thank the KITPC in Beijing for their hospitality and the authors are  thankful to the Erwin Schr\"odinger Institute from Vienna for hospitality while this paper was being written. J.T. is thankful to the Front of Galician-speaking Scientists for encouragement.

This work was supported in part by the ME and FEDER (grant FPA2008- 01838), by the Spanish Consolider-Ingenio 2010 Programme CPAN (CSD2007-00042), by Xunta de Galicia (Conselleria de Educacion and grants PGIDIT06 PXIB206185Pz and INCITE09 206 121 PR). J.T. and J.S. have been supported by ME of Spain under a grant of the FPU program and by the Juan de la Cierva program respectively. J.T. is currently supported by FOM Foundation in Netherlands.

\end{document}